\documentclass[prb,floats,twocolumn,10pt]{revtex4}
\usepackage{graphicx,bm}

\begin{document}

\title {Effects of semiclassical spiral fluctuations on hole dynamics}

\author{I. J. Hamad, L. O. Manuel and A. E. Trumper}

\affiliation {Instituto de F\'{\i}sica Rosario (CONICET) and
Universidad Nacional de Rosario,
Boulevard 27 de Febrero 210 bis, (2000) Rosario, Argentina}

\vspace{4in}
\date{\today}

\begin{abstract}
We investigate the dynamics of a single hole coupled to the spiral fluctuations related to the magnetic ground states of  
the antiferromagnetic $J_1-J_2-J_3$ Heisenberg model on a square lattice. Using exact diagonalization on finite size clusters 
and the self consistent Born approximation in the thermodynamic limit we find, as a general feature, a strong reduction of the 
quasiparticle weight along the spiral phases of the magnetic phase diagram. For an important region of the Brillouin Zone the 
hole spectral functions are completely incoherent, whereas at low energies the spectral weight  is redistributed on several irregular peaks. We find a 
characteristic value of the spiral pitch, ${\bf Q}=(0.7,0.7)\pi$, for which the available phase space for hole scattering is maximum. 
We argue that this behavior is due to the non trivial interference of the magnon assisted and the free hopping mechanism for hole motion, 
characteristic of a hole coupled to semiclassical spiral fluctuations.
\end{abstract}
\maketitle

\section{Introduction}

The interplay between charge and spin degrees of freedom in a 2D doped Mott insulator represents an important problem in condensed matter physics.\cite{Dagotto94,Lee06} The strong constraint on the double occupancy  gives rise to the non trivial coupling  of the charge motion with the magnetic background. Within the context of the $t-J$ model, Shraiman and Siggia\cite{Shraiman89} showed that hole motion produces a long range dipolar distortion of the staggered magnetization resulting in a spiral order characterized by a pitch proportional to hole doping. Further studies focused on the spiral stability under low doping\cite{Sushkov04}  and its possible relation with the spin glass behavior found in the superconducting cuprates. Regarding the charge degrees of freedom, it was suggested that already for the one hole case such a distorsion of the magnetic background would lead to an orthogonality catastrophe,\cite{Anderson90} signalled by the vanishing of the quasiparticle weight $z_{\bf k}$. Subsequent works,\cite{Reiter94} however, showed that such a dipolar distorsion could be compatible with the spin polaron picture where $z_{\bf k}\neq0$ for the whole Brillouin zone (BZ).

In frustrated Mott insulators the single hole dynamics can change drastically. That is the case for the $120^{\circ}$ N\'eel order of the triangular antiferromagnet (AF)\cite{Trumper04}  where the spin polaron picture is not valid for an ample region of the BZ; while for the kagom\'e AF, whose magnetic ground state is disordered,  the hole spectral functions seem to be completely incoherent\cite{Poilblanc04} for the whole BZ. Therefore, the coupling of the hole with the underlying spin excitations is crucial. In the former case, the hole couples with the magnons above the $120^{\circ}$ spiralling magnetization; while in the latter case it couples with an exponentially large number of singlets inside a triplet gap.\cite{Waldtmann98}

A hole can also couple with more exotic magnetic excitations like deconfined spin-$\frac{1}{2}$ spinons.  The deconfined scenario, based on effective fields theories,  was originally  proposed in the presence of short range spiral correlation\cite{Read91} and, later, in the  quantum critical point between N\'eel and valence bond solid phases.\cite{Senthil04} Given that these magnetic correlations are present in the AF $J_1\!-\!J_2\!-\!J_3$ Heisenberg model it is believed that it is an appropriated microscopic model for the search of deconfined spinons.\cite{Read91} Recently, it was hypothesized that by coupling a non magnetic vacancy to the magnetic fluctuations of the above disordered regimes, in particular the plaquette ones, the strong reduction of the vacancy quasiparticle weight  would give a measure of the degree of spinon deconfinement.\cite{Poilblanc06} Even if the idea seems to be appealing the QP weight reduction can be interpreted within a more conventional scenario.\cite{Hamad09}

Very recently, the phase diagram of the ${J_1\!-\!J_2\!-\!J_3}$ model was studied exhaustively  with functional renormalization group, coupled cluster method, and series expansion\cite{Reuther11}. The good agreement among these complementary techniques allowed to establish the location of the quantum disordered (QD) regime in an important region of the parameter space (see right panel of Fig. \ref{fig1}). Interestingly, in this QD regime substantial plaquette and short range incommensurate spiral fluctuations were found, analogously to early exact diagonalization studies on finite systems.\cite{Leung96} For this reason, a non trivial hole dynamics is expected in the QD regime of the above model. 

Motivated by this issue we investigate, as a first step, the spectral function of a hole coupled to the spiral fluctuations of the ${J_1\!-\!J_2\!-\!J_3}$ model. Using the $t\!-\!J$ model, solved with exact diagonalization and within the self consistent Born approximation (SCBA),\cite{Kane89,Martinez92}  we  found that {\it i)} for some regimes the sole inclusion of spiral fluctuations in the SCBA is enough to describe quite well the exact hole spectral functions on finite systems and {\it ii)} that a non trivial hole dynamics is observed under the effect of semiclassical spiral fluctuations in the thermodynamic limit. In particular, we found that in the strong coupling regime ($J/t< 1$) the spin polaron picture is not valid for an ample region of the BZ owing to the non trivial coupling of the hole with the spiral magnetic fluctuations which have been considered within the linear spin wave (LSW) approximation. Notably, in the weak coupling regime, ($J/t\sim10$), where only few magnons are involved, it is observed a strong reduction of the quasiparticle weight averaged on the entire BZ, $Z_{av}\le 0.6$ (see Fig. \ref{fig5}). In general, the QP coherence is lost due to the strong interference of the magnon assisted and the free hopping mechanisms for hole motion of the effective Hamiltonian,  which increases the available phase space for hole scattering. By fine tunning frustration, we found that this effect is maximum when the magnetic pitch is around ${\bf Q}=(0.7\pi,0.7\pi)$. In many cases this spiral pitch corresponds to  quite well ordered spirals, so  we suggest that the QP vanishing is a direct consequence of the hole coupling with semiclassical spiral fluctuations.

The paper is organized as follows: in Sec. IIA we briefly resume the magnetic phase diagram of the $J_1-J_2-J_3$ model along with the LSW calculation. In Sec. IIB we present the $t-J$ model treated within the self consistent Born approximation. In Sec. III  we  compare the spectra predicted by the SCBA  with exact diagonalization on finite clusters and  analyze the spectra in the thermodynamic limit. In Sec IV we close with the conclusions.

\section{\textbf{Models and methods}}

\subsection{$J_1-J_2-J_3$ model}
\begin{figure}[h]
\vspace*{0.cm}
\includegraphics*[width=0.48\textwidth]{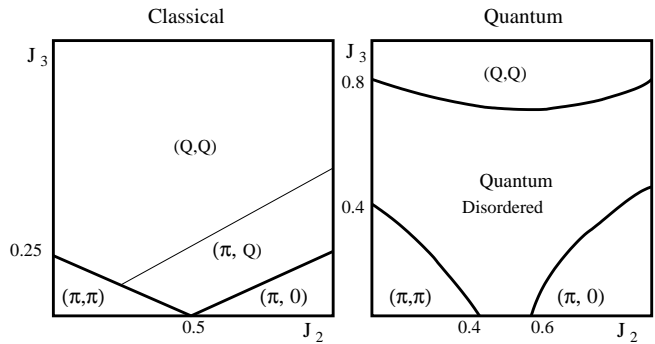}
\caption{Phase diagram of the $\small{J_1\!\!-\!\!J_2\!\!-\!\!J_3}$ model. Left panel is the classical one with the magnetic phases characterized by the spiral pitch indicated  in parenthesis\cite{Ferrer93}.  Right panel: sketch of the quantum phase diagram for $S\!\!=\!\!\frac{1}{2}$ recently found in Ref. \cite{Reuther11}}
\label{fig1}
\end{figure}

The $J_1\!-\!J_2\!-\!J_3$ AF Heisenberg model on a square lattice is defined as:  
\begin{eqnarray}
H_J & = J_1\!\! \displaystyle \sum_{<ij>} \!\!{\bf{S}}_i \cdot {\bf{S}}_j \!\! +\! J_2 \!\!\sum_{<ik>}\!\!{\bf{S}}_{i} \cdot {\bf{S}}_k\!\!+\! J_3 \!\!\sum_{<im>}\!\!{\bf{S}}_i \cdot {\bf{S}}_m ,
\label{JJJ}
\end{eqnarray}

\noindent where ${<\!\!ij\!\!>}$, ${<\!\!ik\!\!>}$, and ${<\!\!im\!\!>}$ indicate sums to first, second, and third neighbors, respectively. The classical phase diagram of this model is shown in the left panel of Fig. \ref{fig1} and contains \textit{N\`eel } $(\pi,\pi)$, \textit{spirals} $(Q,Q)$ and $(Q,\pi)$, and \textit{collinear} $(\pi,0)$ or $(0,\pi)$ phases.\cite{Ferrer93} The thick and thin lines indicate continuous and discontinuous transitions, respectively. Actually, the collinear $(\pi,0)$ and $(0,\pi)$ phases are degenerated with an infinite number of interpenetrating N\'eel states but it is known that quantum fluctuations select the collinear ones by the well known \textit {order by disorder} phenomenon.\cite{Villain77}

 For spin $S\!\!\!=\!\!\!\frac{1}{2}$, a semiclassical spin wave study  predicts an intermediate disordered phase  along the classical transition lines between the classical phases.\cite{Ferrer93} During a long time the quantum $S=\frac{1}{2}$ case of the $J_1-J_2$ model has been intensively studied in the literature.\cite{Richter10} Only recently, however, the complete quantum phase diagram has been investigated  with more sophisticated techniques. In the right panel of Fig. \ref{fig1} it is sketched the quantum phase diagram of the model for $S=\frac{1}{2}$, recently found.\cite{Reuther11}
In particular, in the quantum disordered regime there is evidence of short range plaquette and incommensurate spiral correlations. As we have mentioned in the introduction in the present work we will consider the effects of the spiral fluctuations on the single hole dynamics and leave the effects of plaquette fluctuations for a future work.   
Since the hole dynamics depends mostly on the structure of the magnetic fluctuations, rather than on their long -- or short-- range character, and given that the spiral pitchs practically coincides with the classical ones,\cite{Reuther11} we assume the magnetic background within the linear spin wave theory. For spiral phases the spin operators of $H_J$ (\ref{JJJ}) are  expressed  with respect to a \textit{local axis} pointing in the classical direction of the spin at each site. Using the Holstein-Primakoff transformation in Eq. (\ref{JJJ}), it results the following quadratic Hamiltonian

\begin{equation}
\hat{H}_J =\sum_{\bf q} \gamma_{\bf q} \hat{a}^{\dagger}_{\bf q} \hat{a}_{\bf q} +
\frac{1}{2} \sum_{\bf q} \beta_{\bf q} \left( \hat{a}^{\dagger}_{\bf q}\hat{a}^{\dagger}_{-\bf q}+
\hat{a}_{-\bf q}\hat{a}_{\bf q}\right)+E_{cl}
\label{HP}
\end{equation}

\noindent where

\begin{eqnarray}
\gamma_{\bf q}&=&2s \sum_{\delta>0}J_{\delta} [\cos^2 \frac{{\bf Q}.{\bf \delta}}{2} \cos {\bf q}.{\bf \delta}-\cos {\bf Q}.{\bf \delta}] \nonumber \\ 
\beta_{\bf q}&=&-2s\sum_{\delta>0}J_{\delta} [\sin^2 \frac{{\bf Q}.{\bf \delta}}{2}\cos {\bf q}.{\bf \delta}] \nonumber,
\end{eqnarray}

\noindent being ${\bf Q}$ the spiral pitch that minimizes $ J_{\bf k}=\sum_{\delta} J_{\delta} e^{i {\bf{k.\delta}}}$ with the sums on $\delta$ extending up to third neighbors. After Bogoliubov transformation $\hat{a}_{\bf q}=u_{\bf q}\hat{\alpha}_{\bf q}+v_{\bf q}\hat{\alpha}^{\dagger}_{-\bf q},$
such as 
$u_{\bf q}\!=\![\frac{\gamma_{\bf q}+\omega_{\bf q}}{2\omega_{\bf q}}]^{\frac{1}{2}},$ and
$v_{\bf q}\!=\!-\rm{sign}(\beta_{\bf q}) [\frac{\gamma_{\bf q}-\omega_{\bf q}}{2\omega_{\bf q}}]^{\frac{1}{2}},$ the Hamiltonian is diagonalized as 

$$\hat{H}_J=\sum_{\bf q}\omega_{\bf q}\hat{\alpha}^{\dagger}_{\bf q}\hat{\alpha}_{\bf q}+
\frac{1}{2}\sum_{\bf q}\omega_{\bf q}+(1+\frac{1}{s})E_{cl}.$$

\noindent with a magnon relation dispersion 

\begin{equation}
\small{\omega^2_{\bf q}={s^2 \small{[J_{\bf q}-J_{\bf Q}]}[\frac{1}{2} (J_{{\bf q}+{\bf Q}}+J_{{\bf q}-{\bf Q}})-J_{\bf Q}]}.
=\gamma^2_{\bf q}-\beta^2_{\bf q}}.
\label{dispmag}
\end{equation}

\noindent While the first member is the usual expression used in the literature for the spin wave dispersion,\cite{Ferrer93} the second one, in terms of functions $\gamma_{\bf q}$ and $\beta_{\bf q}$ allows us to write down a more compact expression for the hole-magnon vertex interaction (see below). Notice that besides the three Goldstone modes at ${\bf q}= (0,0),$ and ${\bf Q}=\pm(Q, Q)$, corresponding to the complete $SO(3)$ symmetry rupture, at the linear spin wave level there appear two extra zero modes at ${\bf k}= (-Q,Q)$ and $(Q,-Q)$  that reflect the lattice symmetry in the spectrum (\ref{dispmag}). In fact, a classical spiral  $(Q,Q)$ is related to $(Q,-Q)$ and $(-Q,Q)$ by a global rotation combined with a reflexion about $y$ and $x$ axis, respectively.\cite{Capriotti04} Higher orders beyond linear spin wave theory will lift these degeneracies,\cite{Rastelli92} nevertheless for our purpose  it is enough to keep up to quadratic order since the main effects on the hole dynamics will be related to the non-collinearity of the magnetic background signalled by the spiral pitch ${\bf Q}$ (see below).

\subsection{t-J model and SCBA}
\label{tjmodel}

The dynamics of a hole coupled with the magnetic excitations of a Mott insulator is properly described by the $t-J$ model\cite{Lee06} defined as

\begin{equation}
H_{t-J}=H_t+H_J= -t \sum_{\textless i j \textgreater \sigma} (\tilde{c}^{\dagger}_{i \sigma} \tilde{c}_{j \sigma} + h.c) + H_J  
\end{equation}

\noindent where the electronic operators are the projected ones, 
$\tilde{c}_{i\sigma}=(1-n_{i -\sigma})c_{i\sigma},$ that obey the no double 
occupancy constraint, and in our present case  $H_J$ is Eq. (\ref{JJJ}).
To take care of the constraint we use the spinless fermion representation\cite{Kane89} for $H_t$, leading to the following  effective Hamiltonian for the hole.

\begin{eqnarray}
H_{eff} =  \sum_{k}\epsilon_{k}h_{k}^{\dagger}h_{k}+ \frac{2s}{\sqrt{N}}\sum_{\bf{k,q}} (M_{{\bf{k,q}}} \hat{h}^\dagger_{\bf{k-q}} \hat{h}_{\bf{k}} \alpha^\dagger_{{\bf{q}}}+ h.c.) 
\label{effham}
\end{eqnarray}

\noindent The first term of Eq. (\ref{effham}) describes the free hopping of the hole without disturbing the magnetic background, and is characterized by the hole dispersion $\small{\varepsilon_{{\bf k}}\!=\!2s\sum_{{\bm R}>0} 2t_{{\bm R}}\cos(\frac{\bf{Q \cdot \bm{R}}} {2}) \cos ({{\bf k}}\cdot{{\bm R}})}$ with ${\bm R}\!\!=\!(1,0),(0,1)$, in units of lattice space $a.$ The second term describes the magnon assisted mechanism for the hole motion, and is characterized by the hole-magnon vertex 
$$M_{{\bf k} {\bf q}}=\imath \left(\eta_{{\bf k-q}}u_{{\bf q}}-\eta_{{\bf k}}v_{{\bf q}}\right),
$$

\noindent where $\eta_{{\bf k}}\!=\! 2s  \sum_{{\bm R}\;>0} 2 t_{\bm R} \sin \Large( \frac{{\bf Q} \! \cdot \! {\bm R}}{2} \Large) \sin ({{\bf k}}\cdot{{\bm R}}).$\\

\noindent For a N\'eel phase the free hopping in Eq. (\ref{effham}) vanishes and it is widely accepted the spin polaron picture,\cite{Kane89,Martinez92} where the QP excitations are identified with the coherent propagation of the hole with the magnetic disturbance (magnon cloud). For spiral phases, instead, the free hopping processes interfere with the magnon assisted ones. In particular, we will show that the interference depends strongly on the spiral pitch ${\bf Q}$, which for certain values of frustration leads to the vanishing of $z_{\bf k}$. Regarding  the hole-magnon vertex it is interesting to note that while for the N\'eel phase $M_{{\bf k} {\bf q}}$ vanishes at ${\bf q}=(0,0)$ and $(\pi,\pi),$ for spiral phases it behaves as 

$$
M_{{\bf k} {\bf q}} \simeq -2\imath \left[\frac{\gamma_{\bf 0}}{2\omega_{\bf q}}\right]^{\frac{1}{2}}
\sum_{R>0}t_R \; ({\bf q}\cdot {\bf R})\;\sin \frac{{\bf Q}.{\bf R}}{2}\;\cos {{\bf k}\cdot {\bf R}}
$$

$$M_{{\bf k}, {\bf Q}+{\bf q}}=2\imath \left[\frac{\gamma_{\bf Q}}{2\omega_{{\bf Q}+{\bf q}}}\right]^{\frac{1}{2}}
\sum_{R>0}t_R \sin {\bf Q}.{\bf R}\;\sin ({\bf k}-\frac{\bf Q}{2})\cdot{\bf R}$$

\noindent for small ${\bf q}$. As $\omega_{\bf q}\!\! \sim \!\! |{\bf q}|$ and 
$\omega_{{\bf Q}+{\bf q}}\!\! \sim \!\! |{\bf q}|$, the above vertices behave as 
$M_{{\bf k} {\bf q}} \propto |{\bf q}|^{\frac{1}{2}}$ and $M_{{\bf k}, {\bf Q}+{\bf q}}\propto |{\bf q}|^{-\frac{1}{2}}$ around $(0,0)$ and $(Q,Q)$, respectively. 
To investigate the hole dynamics we compute the hole spectral function  
$A_{\bf k}(\omega)=-(1/\pi)\textrm{Im} G^h_{\bf k}(\omega)$, where 
$G^h_{\bf k}(\omega)=<AF|h_{\bf k}[1/(\omega+i\eta^+-H_{eff})]h^{\dagger}_{\bf k}|AF>$ is
the retarded hole Green function, and $|AF>$ is the undoped magnetic ground state in the
LSW approximation.
In the SCBA, the hole self-energy is given by the following self-consistent equation\cite{Kane89,Martinez92,Trumper04}
$$\Sigma_{\bf k}(\omega)=\sum_{\bf q}\frac{|M_{{\bf k}{\bf q}}|^2}{\omega+i\eta^+-\omega_{\bf q}-
\varepsilon_{{\bf k}-{\bf q}}-\Sigma_{{\bf k}-{\bf q}}(\omega-\omega_{{\bf k}-{\bf q}})}.$$
\noindent which must be solved numerically.
The  QP spectral weight can be calculated as
 $z_{\bf k}=[1-\partial\Sigma_{\bf k}(\omega)/\partial \omega]^{-1}|_{E_{{\bf k}}}$, 
where the QP energy is given by the equation $E_{\bf k}=\epsilon_{\bf k}+Re\Sigma_{\bf k}(E_{\bf k})$.

\section{\textbf{Results: SCBA and Lanczos}}
\label{Results}

\subsection{Finite systems}

In this subsection we compare the hole spectral functions predicted by the SCBA and exact diagonalization on a cluster size of $N=20$ sites. For collinear phases, $(\pi, \pi)$ and $(\pi,0)$, we have already confirmed in a previous work a very  good agreement between both techniques for the hole dynamics in the context of the $J_1-J_2$ model.\cite{Hamad08} Here we are interested in frustration values that induce spiral correlations. In previous exact studies of the $J_1\!\!-J_2\!\!-\!\!J_3$ model\cite{Leung96} it has been shown  a coexistence  of spiral and plaquette fluctuations which, in principle, would couple with the hole. Therefore, a careful comparison of the exact hole spectral functions with that of the SCBA would allow us to discern the relevance, or not, of the spiral fluctuations. It is worth to stress that for finite systems the parameter space is restricted to frustration values that give rise to spiral pitchs coincident with the momenta of the $N=20$ BZ. Even in this case, the strong finite size quantum fluctuations renormalize the classical magnetic wave vector ${\bf Q}_{cl}$. So, in order to perform a faithful comparison, we have assumed the position ${\bf q }_m$ of the maximum exact structure factor $S({\bf q})$ as the actual spiral pitch characterizing the magnetic background within the SCBA. In Fig. 2 it is shown the comparison of the hole spectral functions with Lanczos and SCBA  in the strong coupling regime, $J_1/t=0.4$, along with the exact structure factor for different values of  $J_3/J_1$ at constant values of $J_2$ and hole momentum ${\bf k}_{hole}$. It can be seen that the agreement is reasonably good. In particular, when the magnetic structure factor is characterized by only one main peak like that at $\Delta=(\frac{3}{5}\pi,\frac{4}{5} \pi)$ (see upper panels of Fig. 2) the agreement is quite good, in contrast to the case of two peaks (see $\Delta$ and $X=(\pi,\pi)$ at lower panels of Fig. 2). The latter is related to the strong competence of spiral and N\'eel fluctuations  for this regime; however in the SCBA only the main spiral peak at momentum $\Delta$ is considered. Given that the exact results incorpore all kind of magnetic fluctuations, our results suggest that to describe the main characteristics of the hole spectral function, at least in the regimes considered, it is enough to take into account the coupling of the hole with magnonic excitations above the spiral correlations.
\\

\begin{widetext}
\begin{center}
{\begin{figure}[t]
\vspace*{0.cm}
\includegraphics*[width=0.7\textwidth]{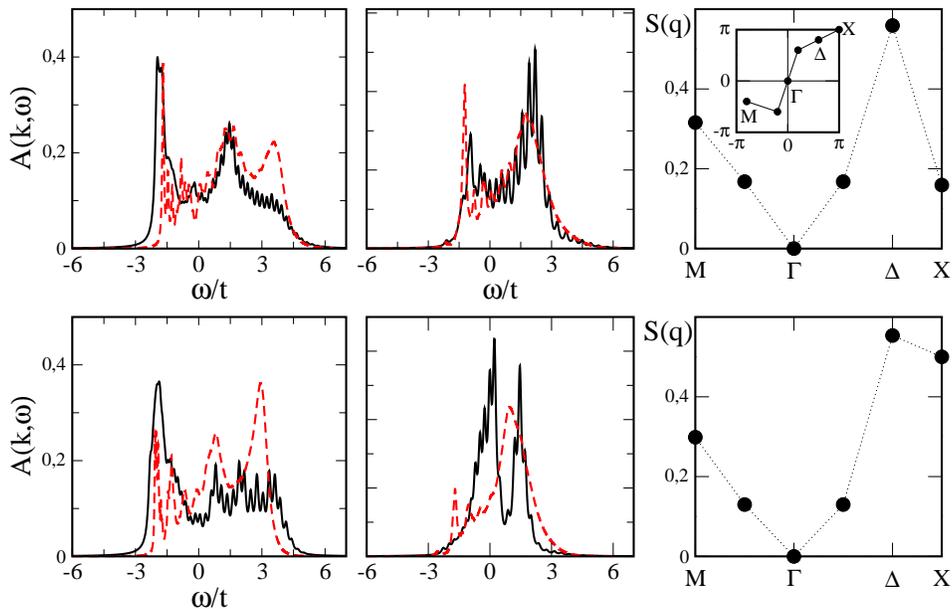}
\caption{Lanczos (continuous) vs. SCBA (dashed) hole spectral functions for a 20 sites cluster, momentum $k_{hole}=(0,\pi)$ (left panels) and $k_{hole}=(0,0)$ (middle panels), and $J_2/J_1=0.2$, $J_1/t=0.4$. Upper (lower) panels correspond to $J_3/J_1=1.5$ ($J_3/J_1=0.45$). Right panels: Exact static structure factor $S({\bf{q}})$. In the SCBA calculation, the magnetic wave vector ${\bf{Q}}$ corresponding to the highest value of $S({\bf{q}})$ in the exact calculation has been chosen. $M=-(0.8,0.4)\pi$, and $\Delta=(0.8,0.6)\pi$.}
\label{fig2}
\end{figure}}
\end{center} 
\end{widetext}

\noindent In addition, in the spiral regime there is a strong reduction of the quasiparticle weight in the low energy sector of the above spectra.This issue will be studied in the next subsection for the thermodynamic limit.

\subsection{Thermodynamic limit}

\begin{figure}[t]
\includegraphics[width=0.45\textwidth]{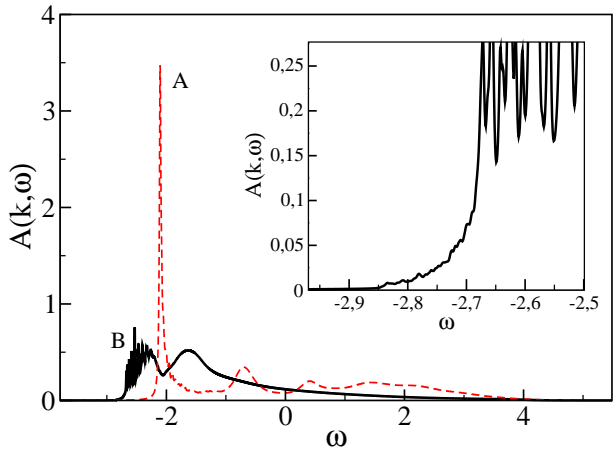}
\caption{Hole spectral functions obtained by SCBA for 1600 sites, $k_{hole}=(0.8,0.8)\pi$, $J_2\!=\!0$, and $J_1/t\!=\!0.4$. Dashed (red) line: $J_3/J_1\!=\!0.1$ (Neel phase). Continuous line: $J_3/J_1\!=\!0.425$ (spiral phase, $Q\!=\!(0.7,0.7)\pi$). Inset:  low energy sector is zoomed. Label A and B, correspond to the points indicated in the magnetic phase diagram of Fig. \ref{fig5} }
\label{fig3}
\end{figure}

\noindent We have computed the hole spectral functions for cluster sizes up to 1600 sites. In complete agreement with previous works we found that for the frustrated N\'eel $(\pi,\pi)$ regime the hole spectral functions show a coherent low energy peak for the whole BZ, which is identified with a well defined quasipaticle excitation, {\it i.e.} spin polaron. This is shown in the dashed line of figure \ref{fig3}  where we have plotted the $A({\bf k},\omega)$ for a hole with momentum ${\bf k}_{hole}\!=\!(0.8\pi,0.8\pi)$   at point $A$  of the  magnetic phase diagram (see of Fig. \ref{fig5}). For this case the QP weight is around  $z_{\bf k}\sim0.2$ which means a considerable contribution of multimagnon processes to the spin polaron wave function. As frustration increases the magnon dispersion bandwidth decreases, allowing to the hole to emit and absorb magnons more easily.\cite{Hamad08} Then, there is an increasing contribution of the multimagnon processes in the spin polaron wave function  accompanied by a reduction of the quasiparticle weight.
Therefore, as quantum fluctuation are enhanced by frustration  the QP weight and the magnetization  decrease monotonically. It can be seen that the QP weight remains finite even when long range order is destroyed, what confirms the idea that the hole dynamic depends mostly on the structure of the magnetic fluctuations rather than on their long, or short, range character.

In the incommensurate spiral phase we have found that the hole spectral functions are completely incoherent for an important region of the BZ. This is shown in the solid line of Fig. \ref{fig3} where it is plotted $A({\bf k},\omega)$ at point $B$ of the magnetic phase diagram (see Fig. \ref{fig5}) for the same ${\bf k}_{hole}$. In the inset of Fig. \ref{fig3}  it is shown the peculiar structure of the low energy sector which is composed by several irregular peaks where no signal of QP excitation is present. In the SCBA this feature of the hole spectral function is typical for an incommensurate spiral correlations and is related to the fact that there are two mechanisms for hole motion --the magnon assisted and the free one--  whose  interference may increase the available phase space for hole scattering, leading to the lost of QP excitations. In fact, this effect depends strongly on the spiral pitch  which can be fine tuned by frustration. For instance, we have found that when the spiral pitch is around ${\bf{Q}}=(0.7,0.7) \pi$ (case $B$ of Fig. \ref{fig3}) the effect is maximum with a vanishing of the QP weight in approximately $50\%$ of the BZ. It is worth to stress that here  the
local magnetization of the spiral state is about $m\sim0.24$ so the non trivial hole dynamics is already present at a semiclassical level. We have found a similar behavior  for a hole injected in other non-collinear magnetic backgrounds such as the $120^{\circ}$ N\'eel order\cite{Trumper04} and  canted N\'eel phases,\cite{Hamad06} although  the incommensurate spiral correlations seems to be  more effective in destroying the QP excitations. It is worth noting that the strong reduction of the QP weight along with the rapid redistribution of the spectral weight on several mulipoles has been observed previously on finite systems. Alternatively,   these features were related to a scenario of spinon deconfinement\cite{Poilblanc06} which, of course, is out of the scope of our present approximation. 

\begin{figure}[t]
\centering
\includegraphics[width=0.45\textwidth]{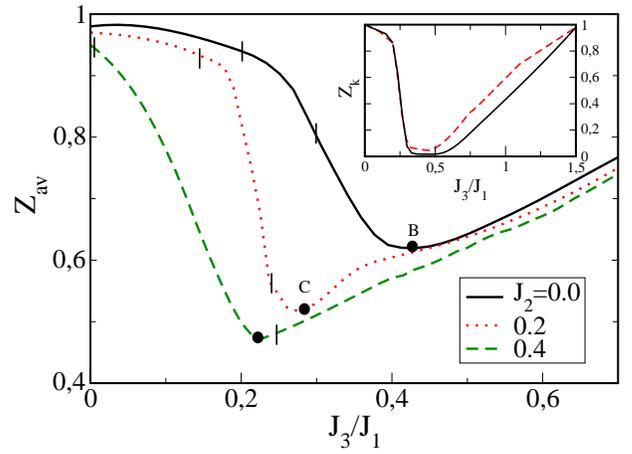}
\caption{QP weight averaged on the entire BZ, $z_{av}=(1/N) \sum_{\bf k} z_{\bf k}$, as a function of $J_3/J_1$, predicted by the SCBA for 1600 sites in the weak coupling regime, $J_1/t=10$. In between the vertical lines $m=0$ within the LSW approximation. The dots $B$ and $C$ indicate the values of frustration for which ${\bf{Q}}\!\approx\!(0.7,0.7) \pi$ (they are displayed in the magnetic phase diagram of Fig. \ref{fig5}. Inset: QP weight for ${\bf k}_{hole}=(0.3,0.3)\pi$ computed with the SCBA (solid line) and Eq. (\ref{zk}) (dashed line).}
\label{fig4}
\end{figure}

We have further investigated the effect of spiral fluctuations for the very weak coupling regime ($J_1/t\gg1$) 
For a N\'eel phase the only mechanism for hole motion is magnon assisted and, owing to its high energy cost,  it is expected that the hole reaches a quasi-static regime where the QP weight approaches unity. For spiral phases, however, the hole can move via the free hopping term of Eq.(\ref{effham}), reducing the QP weight even with a very low average number of magnons promoted. This can be seen from the approximated expression of $z_{\bf k}$ for the weak coupling regime,\cite{Kane89,Hamad08}

\begin{equation}
z_{\bf k}\approx\frac{1}{1+\sum_{\bf q} (M_{{\bf k},{\bf q}}/\epsilon_{\bf k}-\epsilon_{{\bf k}-{\bf q}-\omega_{\bf q}})^2}
\label{zk}
\end{equation}

\noindent In the inset of Fig. \ref{fig4} it is shown the good agreement between the above expression (dashed line) and the QP weight computed with the SCBA (solid line) for different values of frustration. In particular, for the selected hole momentum,  $z_{\bf k}$ is strongly reduced  once the magnetic background reaches the spiral regime. 
In principle, one can attribute the reduced value of $z_{\bf k}$ for spiral phases  to the vertex behavior near $(Q,Q)$, $\small{M_{{\bf k}, {\bf Q}+{\bf q}}\propto {{|{\bf q}}|}^{-\frac{1}{2}}}$. Nonetheless, we have checked that the most important contribution comes from the small values assumed by the denominators $\epsilon_{\bf k}-\epsilon_{{\bf k}-{\bf q}}-\omega_{\bf q}$.  
In order to quantify globally this behavior  we have plotted  in Fig. \ref{fig4}  the QP spectral weight averaged on the entire BZ,  $z_{av}\!\!=\!\!(1/N)\sum_{\bf k} z_{\bf k},$ for $J_1/t=10$. Again,
it can be observed that as $J_3/J_1$ increases $z_{av}$ decreases notably when the magnetic background reaches the spiral regime. In particular, for $J_2/J_1<0.5$ it is always observed a single minimum at ${\bf{Q}}\approx(0.7,0.7)\pi$, analogously to that found in the strong coupling regime  (Fig. \ref{fig3}). For $J_2/J_1=0$ (continuous line in Fig. \ref{fig4}), ${\bf Q}$ occurs in a region where $m\approx0.24$, while for $J_2/J_1=0.4$ (dashed line) ${\bf Q}$ is inside the region where $m=0$. In the latter we have modified the LSW approximation by introducing a gap in the magnetic dispersion in order to describe the short range spiral order properly.\cite{Takahashi87}

\begin{figure}[t]
\includegraphics[width=0.45\textwidth]{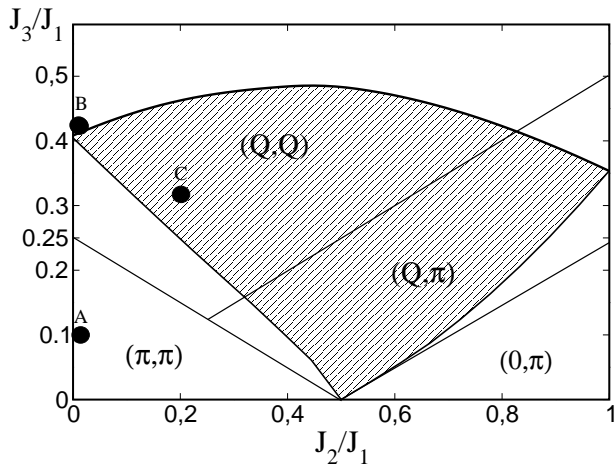}
\caption{The shaded area indicates the region of the magnetic phase space where the averaged quasiparticle weight $z_{av}\!\!<\!\! 0.6$ in the weak coupling regime, ($J_1/t=10$). For the strong coupling regime, $J_1/t=0.4$, the SCBA predicts the vanishing of $z_{\bf k}$ in some area of the BZ. $A=(0,0.1)$, $B=(0,0.425)$, and $C=(0.2,0.32)$.}
\label{fig5}
\end{figure}

Finally, in Fig. \ref{fig5} we plotted the approximated area of the magnetic phase diagram where the QP weight is strongly reduced, that is,  $z_{AV}<0.6$ for the very weak coupling regime. The fact that the shaded area is located at the spiral region of the magnetic phase diagram is a clear indication of the crucial effect of the spiral fluctuation on  the  hole dynamics. Since this is independent of the range of the magnetic fluctuations we expect the same effect for a hole coupled to the short range spiral fluctuation of the actual quantum magnetic phase diagram (see right panel of Fig. \ref{fig1}).

\section{\textbf{Conclusions}}
\label{conslusions}

We have investigated the effect of semiclassical spiral fluctuations 
on a single hole dynamics. Based on the magnetic phase diagram of the $J_1\!\!-\!\!J_2\!\!-\!\!J_3$ model and solving the hole Green function with exact diagonalization and within the SCBA, we found that for the weak  ($J_1/t\gg1$) and the strong ($J_1/t\leq1$) coupling regime there is characteristic value of the spiral pitch --${\bf Q}\!\!\approx\!\!(0.7,0.7)\pi$-- for which the available phase space for hole scattering is maximum. Notably, for the whole spiral regimes ($(Q,Q)$ and $(Q,\pi)$)  of the model we found a strong reduction of the averaged QP weight $z_{AV}$. In particular, for some momenta the QP weight vanishes and the spectral weight at low energy is redistributed on several irregular peaks, or multipoles. Even if in our study the spirals have been described semiclassically we think that this effect should be also observed in the short range spirals phases recently found in the disordered regime of the quantum phase diagram.\cite{Reuther11} Similar features have been found in finite size studies of the same model, although the strong reduction of the QP weight have been attributed to the spinon deconfinement inherent of the plaquette fluctuations.\cite{Poilblanc06}  Based on our results and given that the alternative scenario for spinon deconfinement is based on short range spiral correlations,\cite{Read91} it would be important to go beyond  the semiclassical description and investigate, within the context of the present model, the hole dynamics on spirals treated in terms of spinon excitations.\cite{Takei04} Work in this direction is in progress. 

We thanks A. E. Feiguin for his valuable help and useful discussions. This work was supported by CONICET under Grant PIP2009 Nº 1948.

\end{document}